# First Measurements of Nuclear Detonation Debris with Decay Energy Spectroscopy

Mark P. Croce, Katrina E. Koehler, Veronika Mocko, Andrew S. Hoover, and Stosh A. Kozimor
Los Alamos National Laboratory, Los Alamos, NM 87545

Daniel R. Schmidt and Joel N. Ullom
National Institute of Standards and Technology, Boulder, CO 80305

**Abstract**: We report the first isotopic composition measurements of trinitite, nuclear detonation debris from the Trinity test, using the novel forensics technique of decay energy spectroscopy (DES). DES measures the unique total decay energy (Q value) of each alpha-decaying isotope in a small radioactive sample embedded in a microcalorimeter detector. We find that DES can measure the major alpha-decaying isotopes in small particles of trinitite with no dissolution or chemical processing. These first measurements demonstrate the potential of DES to provide a radiometric isotopic characterization method with sensitivity and precision to complement traditional forensics techniques.

## Introduction

Determining the plutonium isotopic composition of nuclear detonation debris is an important forensics goal. We present a novel methodology, decay energy spectroscopy, for this analysis to provide a complementary technique to more traditional analyses. We have applied this method to a sample of trinitite.

Traditional forensics methodologies usually include alpha spectroscopy and mass spectrometry. Each of these approaches has advantages and disadvantages, leading to a combination of approaches for a thorough forensics analysis.

In standard alpha spectroscopy with energy resolution of 15–20 keV, the 5.486 MeV alpha from $^{241}$Am cannot be resolved from the 5.499 MeV alpha from $^{238}$Pu. Because of interferences like this, chemical separation is done before source preparation in order to resolve the two radioisotopes. To avoid energy loss within the sample, alpha sources must be thin and uniform, which often means electrodeposition if the sample size is small.[1]

Mass spectrometry yields high-precision and high-accuracy isotopic ratios. Mass spectrometry works by ionizing a small sample and then using their mass-to-charge ratio to separate the ions. This method typically requires chemical separations to avoid the isobaric interferences, for example, between $^{241}$Am and $^{241}$Pu.[1]

The radiochemical preparation for both alpha spectroscopy and mass spectrometry can be time consuming, delaying results in a timely forensics investigation. Further, the chemical process serves to homogenize the sample, averaging out any spatial information within the sample.[2–4] As a result, complementary forensics measurements that do not require chemical separation and utilize smaller sample sizes could be valuable.

## Forensic Analyses of Trinitite for Pu Isotopics

The most cited measurement on trinitite by Parekh et al.[5] was done using a combination of alpha and gamma spectroscopy to determine the isotopic composition. In the radiochemical preparation, the aggregate 5.4 g of trinitite was milled and then a 0.5 g aliquot dissolved in $HNO_3$ and HF overnight. The solution was evaporated until dry, and the dissolution process was repeated three times before a final treatment with $HNO_3$. The actinides were then scavenged from the solution using iron hydroxide, separated and purified using ion-exchange chromatography before being electroplated for alpha spectroscopy. The achieved $^{240}$Pu/$^{239}$Pu mass ratio normalized to the time of explosion was 0.013 ± 0.003.

Alpha spectroscopy was used by Belloni et al.[6] to determine the plutonium isotopics. No radiochemical separations were done for this measurement, but due to the energy resolution of alpha spectroscopy with conventional detectors, only a $^{239+240}$Pu/$^{241}$Am ratio could be determined.

Bellucci et al.[4] analyzed the uranium isotopic composition of trinitite with laser ablation multicollector inductively coupled plasma mass spectrometry and inferred a $^{240}$Pu/$^{239}$Pu mass ratio of 0.01–0.03. This technique also differs from traditional mass spectroscopy because it does not require radiochemical separation, so it is able to determine the degree of inhomogeneity within the sample.

A recent radiochemical analysis in this issue reported by Hanson and Oldham[7] gives a $^{240}$Pu/$^{239}$Pu mass ratio of 0.0246(3). This is the most precise trinitite isotopic measurement to date.

**Decay Energy Spectroscopy as a Forensic Tool**

High-resolution calorimetric decay energy spectroscopy (DES) is a recently developed radiometric technique in which a radioactive sample is embedded within the absorber of a low-temperature microcalorimeter.[8,9] For each nuclear decay, the energy of all decay products (alpha particles, gamma-rays, X-rays, electrons, etc.) heats the absorber. Therefore, energy loss in the sample is less of a factor in DES than in alpha spectroscopy, and sample preparation can be simplified. The detector measures 100% of alpha decays, and so the sensitivity of DES is the highest achievable by any radiometric method. Because of the small sample sizes used, separate measurements can be made on sub-samples, a technique which reduces the homogenizing effect of measuring large samples. The temperature change of the absorber is measured by a very sensitive thermometer such as a superconducting transition-edge sensor. For alpha-decaying nuclides ($^{239}$Pu, $^{240}$Pu, $^{241}$Am, etc.), the measured energy corresponds to the unique total nuclear decay energy (Q value). Operation at low temperatures (below 0.1 K) is achieved with commercial cryostat systems and enables ultra-high energy resolution, as good as 1.0 keV FWHM at 5.5 MeV.[9] These characteristics make DES a very sensitive method complementary to traditional forensics techniques for robust, precise isotopic analysis. In this paper, we report the first DES measurements of nuclear detonation debris.

**Decay Energy Spectroscopy Measurements of Trinitite Particles**

DES has been successfully applied to undissolved pure plutonium oxide particles collected from a certified reference material.[9] Trinitite, debris from the first nuclear detonation, contains only small quantities of plutonium dispersed through a complex matrix and represents a more challenging measurement. To produce small particles for measurement by DES, a 9.16 mg aerodynamic bead (Figure 1A) was ball milled (dry) for 10 minutes. Approximately 50 μg of the resulting powder was embedded in a gold foil by repeatedly folding and pressing the gold foil with pliers to form an absorber for the microcalorimeter detector. A cross section of a representative gold sample with embedded trinitite particles is shown in Figure 1B. The size of trinitite particles varied significantly, but nearly all were below 10 μm in diameter. The assembled microcalorimeter detector is shown in Figure 1C, with the gold absorber attached to the transition-edge sensor chip by an indium bond. The acquired DES spectrum of trinitite shows the presence of beta-decaying fission and neutron activation products, along with alpha-decaying actinides (Figure 2). The acquisition time was 19.4 hours, and the total measured alpha activity was 0.04 Bq. A second-order polynomial was used for energy calibration, based on the $^{238}$Pu and $^{239}$Pu peaks. Beta-decaying nuclides cannot be individually quantified, but their aggregate presence is a signature of nuclear fission. The energy resolution is sufficient to identify the presence of $^{239}$Pu, $^{240}$Pu, $^{238}$Pu, and $^{241}$Am, but degraded relative to DES measurements of pure plutonium materials (Figure 3).

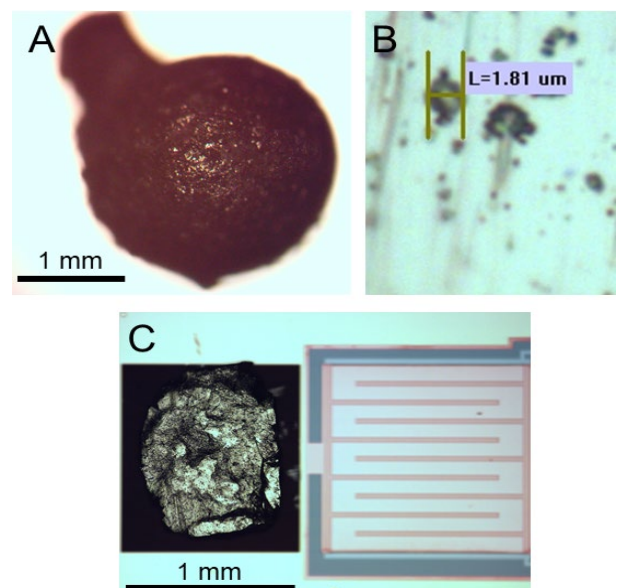

Figure 1. (A) Aerodynamic bead of trinitite used for this measurement. (B) Cross section of representative gold sample shows ball-milled trinitite particles were effectively embedded with a simple folding and pressing process. (C) The gold absorber with embedded particles was attached to the transition-edge sensor with an indium bond. Sample preparation required no dissolution or chemical processing.

Figure 3 (top) shows the $^{239}$Pu and $^{240}$Pu peaks with a preliminary fit to the data. This region of the spectrum was fit with two exponential-tailed Gaussian functions, constrained to have the same Gaussian width and tail factor, and a difference in Gaussian centroid values equal to the known difference in Q values between $^{239}$Pu and $^{240}$Pu. The Gaussian component FWHM is 5.1 keV, and the exponential tail factor is 21 keV. Previous DES measurements on pure materials have demonstrated a Gaussian FWHM of 1.0 keV with negligible tailing.[8] We expect that the energy resolution and tailing could be improved by further developing the milling process to produce smaller particles, such that more decay energy would be deposited in the gold absorber matrix rather than the particles of trinitite.

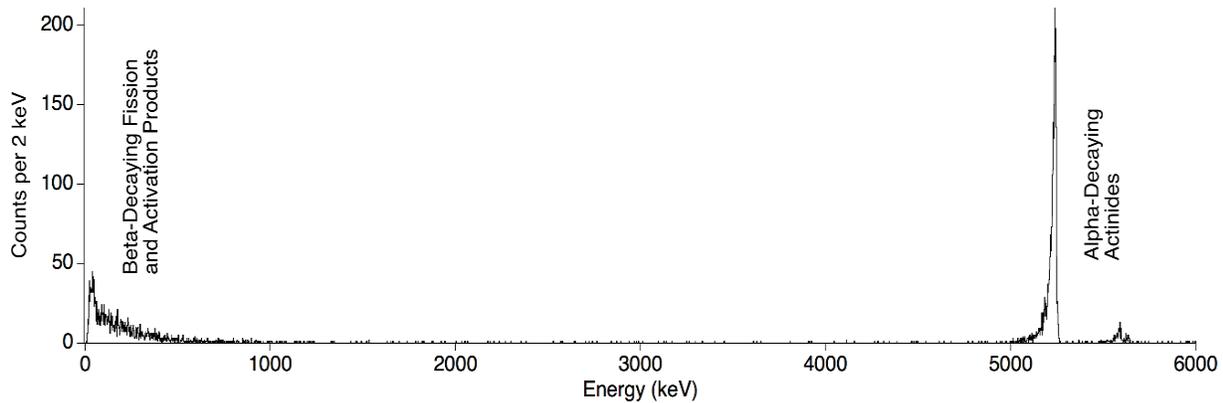

Figure 2. The full DES spectrum of trinitite indicates beta-decaying fission and neutron activation products, primarily below 1 MeV, and alpha-decaying actinides above 5 MeV.

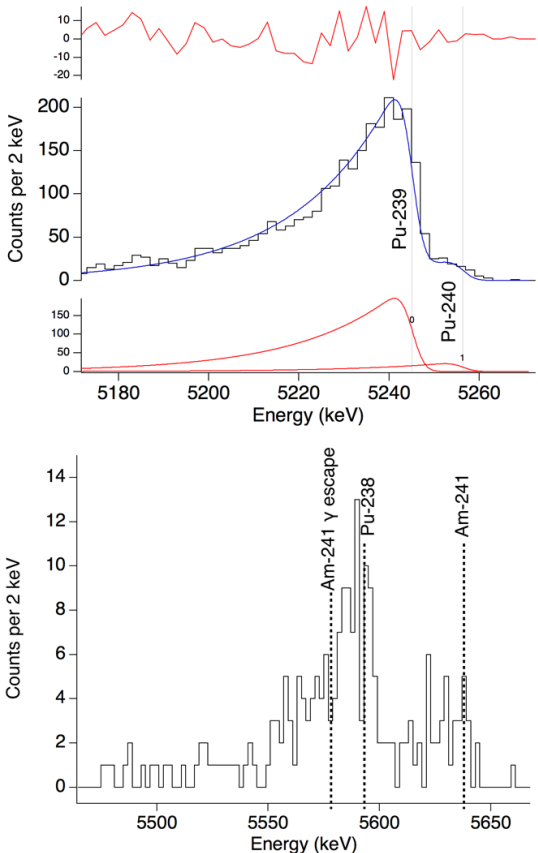

Figure 3. (Top) Peaks from $^{239}$Pu and $^{240}$Pu are observed in trinitite. The middle plot shows the fitted spectrum, the top plot shows residuals, and the bottom plot shows the fit components. (Bottom) The presence of $^{238}$Pu and $^{241}$Am can also be identified. Dashed lines indicate the expected peak locations. A peak from the $^{241}$Am 59.5 keV gamma ray escape is expected and has been observed in other measurements.

A preliminary analysis indicates approximately 16 pg of $^{239}$Pu and approximately 0.5 pg of $^{240}$Pu. While further work is required to quantify the uncertainty of this result, the decay-corrected $^{240}$Pu/$^{239}$Pu mass ratio of 0.03 at the time of explosion is consistent with the results of Bellucci et al. who reported mass ratios ranging from 0.01–0.03.[4] Our mass ratio is higher than the value reported by Parekh *et al.*[5] (0.013). It is closer to the value of Hanson and Oldham[7] (0.0246).

Figure 3 (bottom) shows the region of the spectrum with peaks from $^{238}$Pu and $^{241}$Am. In the measured spectrum, the activity of these two isotopes is too low to merit quantitative analysis, but their presence can be identified. With an intrinsically low background, DES uncertainty is primarily limited by counting statistics and could be improved with a longer measurement time. It may also be possible to load more trinitite particles into the gold matrix and increase the activity of the sample. It is important to note that there was no cleanup involved in the sample preparation. This is an extremely impure sample, measured with no chemical processing or even dissolution. The plutonium concentration in similar trinitite samples has been measured at less than 1 ppm by mass. This first measurement of its type therefore shows the potential of DES to measure actinide isotopic composition in extremely impure materials with minimal sample preparation.

**Conclusions**

These first decay energy spectroscopy measurements of nuclear detonation debris demonstrate the potential value of the technique. DES was able to identify $^{238}$Pu, $^{239}$Pu, $^{240}$Pu, and $^{241}$Am and obtain a decay-corrected $^{240}$Pu/$^{239}$Pu mass ratio of 0.03 in a trinitite bead with a simple, rapid sample preparation process and no dissolution or chemical processing on a sample three orders of magnitude smaller than the sample size used in the Parekh et al. analysis. With further development, we anticipate that the precision and sensitivity of DES can make it a valuable complementary method to mass spectrometry techniques to resolve interferences, provide a timely response on actinide isotopic composition without requiring time-consuming chemical separations, and provide information on the inhomogeneity of various samples by measuring rather small samples.

## Acknowledgements
This work was supported by the Los Alamos National Laboratory Pathfinder Program. The microcalorimeter sensors for decay energy spectroscopy were designed and fabricated by the National Institute of Standards and Technology Quantum Sensors Group.